\documentclass[11pt,a4paper,twoside]{JHEP3}

\usepackage{cite}
\usepackage{graphicx}
\usepackage{epsfig}
\usepackage{dcolumn}
\usepackage{bm}

\def \d {{\rm d}}

\def \e {{\epsilon}}
\def \P {{p_\lambda}}
\def \Q {{p_\nu}}

\newcommand{\be}{\begin{equation}}
\newcommand{\ee}{\end{equation}}

\newcommand{\beqn}{\begin{eqnarray}}
\newcommand{\eeqn}{\end{eqnarray}}

\newcommand{\AS}{Aichelburg-Sexl }
\newcommand{\pa}{\partial}
\newcommand{\pp}{{\it pp\,}-}
\newcommand{\ba}{\begin{array}}
\newcommand{\ea}{\end{array}}

\title{Ultrarelativistic boost of spinning black rings}

\author{Marcello Ortaggio,\thanks{also at INFN, Rome, Italy}\, Ji\v{r}\'{\i} Podolsk\'y and Pavel Krtou\v{s} \\ 
        Institute of Theoretical Physics, Faculty of Mathematics and Physics, \\ Charles University in Prague,
  V Hole\v{s}ovi\v{c}k\'{a}ch 2, 180 00 Prague 8, Czech Republic \\
        {E-mail: \email{marcello.ortaggio@comune.re.it}, \email{Jiri.Podolsky@mff.cuni.cz}, \email{Pavel.Krtous@mff.cuni.cz}}}

\abstract{We study the ${D=5}$ Emparan-Reall spinning black ring under an ultrarelativistic boost along an arbitrary direction. We analytically determine the resulting shock \pp wave, in particular for boosts along axes orthogonal and parallel to the plane of rotation. The solution becomes physically more interesting and simpler if one enforces equilibrium between the forces on the ring. We also comment on the ultrarelativistic limit of recently found supersymmetric black rings with two independent angular momenta. Essential distinct features with respect to the boosted Myers-Perry black holes are pointed out.} 


\begin{document}

\section{Introduction}
\label{sec-introduction}

Shock \pp wave geometries describe the spacetime surrounding very fast moving objects, and are thus relevant to the study of Planckian scattering \cite{tHooft87}. They are also of interest in string theory, since strings may be exactly solved in such backgrounds \cite{AmaKli88,devSan89}. The prototype of shock wave solutions is the \AS spacetime, which represents the gravitational field of a massless point particle. It was originally obtained by boosting the Schwarzschild black hole to the speed of light, while rescaling the mass to zero in an appropriate way \cite{AicSex71}. According to recent extra-dimension scenarios, the fundamental Planck scale of (higher dimensional) gravity could be as low as a few TeV. This has stimulated renewed interest in the study of gravitational effects in high energy collisions, especially in view of the possible observation of microscopic black holes at near future colliders \cite{BanFis99,EmpHorMye00prl,GidTho02,DimLan01} (see, e.g., \cite{CarBerCav05} for a recent review and for further references). It has been shown that closed trapped surfaces do indeed form in the ultrarelativistic collision of \AS point particles \cite{EarGid02,YosNam02} and of finite-size beams \cite{KohVen02}, which can more accurately model string-size effects. Nevertheless, it is desirable to understand how other effects could influence high energy scattering. A first step in this direction is to investigate more general shock wave solutions of higher dimensional gravity, which can naturally be obtained by applying the boosting technique of \cite{AicSex71} to black hole spacetimes. This has been done in any $D\ge 4$ for static black holes with electric charge \cite{LouSan90} or immersed in an external magnetic field \cite{Ortaggio04,Ortaggio05}. The ultrarelativistic limit of the Myers-Perry rotating black holes \cite{MyePer86} has been studied in \cite{Yoshino05}  (for the case of one non-vanishing spin). However, a striking feature of General Relativity in $D>4$ is the non-uniqueness of the spherical black holes of \cite{MyePer86}. In five-dimensional vacuum gravity, there exist also asymptotically flat rotating black rings with an event horizon of topology $S^1\times S^2$ \cite{EmpRea02prl}. In the present contribution, we aim at studying the gravitational field generated by such rings in the \AS limit. As we will see in detail, this results in shock waves generated by extended lightlike sources (with a characteristic length-scale) which are remnants of the ring singularity of the original spacetime \cite{EmpRea02prl}. Our recent results on boosted non-rotating black rings \cite{OrtKrtPod05} will be recovered as a special subcase. In general, the presence of spin is important because it allows black rings to be in equilibrium \cite{EmpRea02prl} without introducing ``unphysical'' membranes via conical singularities \cite{EmpRea02prd}. This will be reflected also in the shock geometry resulting from the boost. 
From a supergravity and string theory point of view, it is remarkable that supersymmetric black rings have been also constructed \cite{Elvangetal04,Elvangetal05,BenWar04,GauGut05}. We will conclude this article with a brief comment on the boost of such solutions. In the Appendix, we compare our results with those obtained for the ultrarelativistic limit of Myers-Perry black holes \cite{Yoshino05} in $D=5$.

\section{The black ring solution}

\label{sec-ring}

In this section we briefly summarize the basic properties of the black ring, referring to \cite{EmpRea02prl,Emparan04} for details. In the coordinates of \cite{Emparan04},\footnote{Up to simple constant rescalings of $F(\zeta)$, $G(\zeta)$, $C(\lambda,\nu)$, $\psi$ and $\phi$, cf.~Eqs.~(\ref{FG}) and (\ref{range}) with the corresponding ones in \cite{Emparan04}. In addition, multiply our $L^2$ by $(1-\nu)/(1-\lambda)$ to obtain the parameter used in \cite{Emparan04}.} the line element reads
\beqn
 \d s^2= & & -\frac{F(y)}{F(x)}\left(\d t+C(\nu,\lambda)L\frac{1+y}{F(y)}\d\psi\right)^2 \nonumber \label{ring} \\
 & & {}+\frac{L^2}{(x-y)^2}F(x)\left[-\frac{G(y)}{F(y)}\d\psi^2-\frac{\d y^2}{G(y)}+\frac{\d x^2}{G(x)}+\frac{G(x)}{
      F(x)}\d\phi^2\right] ,
\eeqn
where
\be
 F(\zeta)=\frac{1+\lambda\zeta}{1-\lambda} , \qquad G(\zeta)=(1-\zeta^2)\frac{1+\nu\zeta}{1-\nu} , \qquad    
 C(\nu,\lambda)=\sqrt{\frac{\lambda(\lambda-\nu)(1+\lambda)}{(1-\nu)(1-\lambda)^3}} .
 \label{FG}
\ee
The dimensionless parameters $\lambda$ and $\nu$ satisfy $0\le\nu\le\lambda<1$, and for $\lambda=0=\nu$ the spacetime (\ref{ring}) is flat. The constant $L>0$ represents a length related to the radius of the ``central circle'' of the ring. For a physical interpretation of the spacetime (\ref{ring}) we take $y\in(-\infty,-1]$, $x\in[-1,+1]$ (see a discussion in \cite{PraPra05} for other possible choices) and $\psi$ and $\phi$ as periodic angular coordinates (see below). Surfaces of constant $y$ have topology $S^1\times S^2$. The coordinate $\psi$ runs along the $S^1$ factor, whereas $(x,\phi)$ parametrize $S^2$ (see \cite{EmpRea02prd,Emparan04,Elvangetal05} for illustrative pictures). Within the above range, $y$ parametrizes ``distances'' from the ring circle. At $y\to-\infty$ the spacetime has a inner spacelike curvature singularity, $y=-1/\nu$ is a horizon and $y=-1/\lambda$ an ergosurface, both with topology $S^1\times S^2$. The black ring solution~(\ref{ring}) is asymptotically flat near spatial infinity $x,y\to -1$, where it tends to Minkowski spacetime in the form
\be
 \d s^2_0=-\d t^2+\frac{L^2}{(x-y)^2}\left[(y^2-1)\d\psi^2+\frac{\d y^2}{y^2-1}+
    \frac{\d x^2}{1-x^2}+(1-x^2)\d\phi^2\right] .
  \label{background}  
\ee
To avoid conical singularities at the axes $x=-1$ and $y=-1$, the angular coordinates must have the standard periodicity 
\be
 \Delta\phi=2\pi=\Delta\psi .
 \label{range}
\ee
Centrifugal repulsion and gravitational self-attraction of the ring are in balance if conical singularities are absent also at $x=+1$, which requires 
\be
 \lambda=\frac{2\nu}{1+\nu^2} .
 \label{equilibrium}
\ee
When this equilibrium condition holds, the metric (\ref{ring}) is a vacuum solution (of $D=5$ General Relativity) everywhere. With different choices (e.g., in the static limit $\nu=\lambda$ \cite{EmpRea02prd}), the conical singularity at $x=+1$ describes a disk-shaped membrane inside the ring. 

The mass, angular momentum and angular velocity (at the horizon) of the black ring are 
\be
 M=\frac{3\pi L^2}{4}\frac{\lambda}{1-\lambda} , \qquad   
 J=\frac{\pi L^3}{2}\sqrt{\frac{\lambda(\lambda-\nu)(1+\lambda)}{(1-\nu)(1-\lambda)^3}} , \qquad   
 \Omega=\frac{1}{L}\sqrt{\frac{(\lambda-\nu)(1-\lambda)}{\lambda(1+\lambda)(1-\nu)}} .
 \label{mass}
\ee
The algebraic type of the Weyl tensor of the ring spacetime is $I_i$ \cite{PraPra05}.

\section{General boost}

\label{sec-general}

For our purposes, it is convenient to decompose the line element (\ref{ring})
as
\begin{equation}
 \d s^2=\d s_0^2+\Delta ,
  \label{decomposition}
\end{equation}
in which $\d s^2_0$ is Minkowski spacetime~(\ref{background}) and
\beqn
 \Delta= & & \lambda\frac{x-y}{1+\lambda x}\d t^2-2(1-\lambda)C(\lambda,\nu)L\frac{1+y}{1+\lambda x}\d t\,\d\psi \nonumber \label{perturbation} \\
  & & {}+\frac{\lambda-\nu}{1-\nu}\frac{L^2}{1+\lambda y}\left[-\lambda\frac{1+\lambda}{1-\lambda}\frac{(1+y)^2}{1+\lambda x}+\frac{y^2-1}{x-y}\right]\d\psi^2 \nonumber \\  & &
{}+\frac{L^2}{(x-y)^2}\left[\nu\frac{x+1}{1-\nu}(y^2-1)\d\psi^2+\frac{\lambda(1-\nu)(x-y)+(\lambda-\nu)(1+y)}{(1-\lambda)(1+\nu y)}\frac{\d y^2}{y^2-1}\right. \nonumber \\
 & & {}\hspace{2cm}+\left.\frac{\lambda-\nu}{1-\lambda}\frac{\d x^2}{(1-x)(1+\nu x)}+\nu\frac{x+1}{1-\nu}(1-x^2)\d\phi^2\right] .
\eeqn The above splitting is such that near infinity ($x,y\to -1$) one has $\d
s^2\to\d s_0^2$, while $\Delta$ becomes ``negligible'' (in the sense of the
``background'' metric $\d s_0^2$). This enables us to define a notion of
Lorentz boost using the symmetries of the asymptotic Minkowskian background $\d
s_0^2$. Cartesian coordinates will visualize it most naturally. These can be
introduced in two steps. First, we replace the coordinates $(y,x)$ with new
coordinates $(\xi,\eta)$ via the substitution \be
 y=-\frac{\xi^2+\eta^2+L^2}{\Sigma} \, , \qquad
 x=-\frac{\xi^2+\eta^2-L^2}{\Sigma} \, ,
 \label{cylindrical}
\ee where \be
 \Sigma=\sqrt{(\eta^2+\xi^2-L^2)^2+4L^2\eta^2} .
 \label{Sigma}
\ee The flat term $\d s_0^2$ in Eq.~(\ref{decomposition}) now takes the form
$\d s_0^2=-\d t^2+\d\eta^2+\eta^2\d\phi^2+\d\xi^2+\xi^2\d\psi^2$. Then,
Cartesian coordinates adapted to the Killing vectors $\pa_\phi$ and $\pa_\psi$
are given by
\be
 x_1=\eta\cos\phi , \quad x_2=\eta\sin\phi , \qquad \qquad y_1=\xi\cos\psi , \quad y_2=\xi\sin\psi ,
 \label{cartesian}
\ee so that $\eta=\sqrt{x_1^2+x_2^2}$, $\xi=\sqrt{y_1^2+y_2^2}$, and $\d
s_0^2=-\d t^2+\d x_1^2+\d x_2^2+\d y_1^2+\d y_2^2$. This enables us to study a
boost along a general direction. Since the original spacetime (\ref{ring}) is
symmetric under (separate) rotations in the $(x_1,x_2)$ and $(y_1,y_2)$ planes,
such a direction can be specified by a single parameter~$\alpha$, namely
introducing rotated axes $z_1$ and $z_2$ \be
 x_1=z_1\cos\alpha-z_2\sin\alpha , \qquad y_1=z_1\sin\alpha+z_2\cos\alpha .
 \label{rotated}
\ee Defining now suitable double null coordinates $(u',v')$ by
\begin{equation}
 t=\frac{-u'+v'}{\sqrt{2}} , \qquad z_1=\frac{u'+v'}{\sqrt{2}} ,
 \label{nullcoords}
\end{equation}
a Lorentz boost along $z_1$ takes the simple form
\begin{equation}
 u'=\epsilon^{-1}u  , \qquad v'=\epsilon v  .
 \label{lorentzboost}
\end{equation}
The parameter $\epsilon>0$ is related to the standard Lorentz factor via
$\gamma=(\epsilon+\epsilon^{-1})/2$. We are interested in ``ultrarelativistic''
boosts to the speed of light, i.e. in taking the limit $\e\to 0$ in the
transformation~(\ref{lorentzboost}). While $\e\to 0$, we will rescale the mass
as $M=\gamma^{-1}p_M\approx2\e p_M$ \cite{AicSex71}, which physically means that the total energy remains finite in the limit ($p_M>0$ is a constant). Moreover, during the ultrarelativistic limit we wish
to keep the angular velocity $\Omega$ finite (a similar condition was imposed
in \cite{Yoshino05}), and to allow for the possibility of black rings in
equilibrium [when the condition~(\ref{equilibrium}) holds]. From
Eq.~(\ref{mass}), these requirements imply the rescalings\footnote{This appears
to be physically the most interesting and simple choice. See Footnote~\ref{note-rescaling} for a subtler, slightly more general comment.} 
\be
 \lambda=\epsilon \P , \qquad  \nu=\epsilon \Q ,
 \label{ASlambda}
\ee where $p_\lambda=8p_M/(3\pi L^2)$ and $p_\nu$ is another positive constant
such that $\P\ge\Q$. In terms of these parameters, for $\e\to 0$ the
equilibrium condition~(\ref{equilibrium}) becomes \be
 \P=2\Q .
 \label{equilibrium2}
\ee Values ${\Q\le\P<2\Q}$ correspond to black rings~(\ref{ring}) which are
``underspinning'' before the boost (and therefore balanced by a membrane of negative
energy density), values $\P>2\Q$ to ``overspinning'' black rings (with a
membrane of positive energy density). Notice, however, that under the limit
$\e\to 0$ the angular momentum $J$ will tend to zero (as~$\sim\e$).

We can now evaluate how the black ring metric~(\ref{ring}) [that is,
Eq.~(\ref{decomposition}) with Eqs.~(\ref{background}) and
(\ref{perturbation})] transforms under the boost~(\ref{lorentzboost}). We have
first to substitute Eq.~(\ref{cylindrical}) into Eqs.~(\ref{background}) and
(\ref{perturbation}). Then, we apply the sequence of
substitutions~(\ref{cartesian}), (\ref{rotated}), (\ref{nullcoords}) into the
thus obtained expressions for $\d s_0^2$ and for $\Delta$. Finally, we
perform the boost~(\ref{lorentzboost}) with the rescalings~(\ref{ASlambda}), which make $\Delta=\Delta_\e$ dependent on $\e$. The $\d s_0^2$ is invariant under the boost and at the end it reads
\be
 \d s_0^2=2\d u\d v+\d x_2^2+\d y_2^2+\d z_2^2 .
 \label{back}
\ee 
The next step is to take the ultrarelativistic limit ${\d s^2=\d s_0^2+\lim_{\e\to 0}\Delta_\e}$. This is delicate because the expansion of ${\Delta_\e}$ in ${\e}$
has a different structure in different regions of the spacetime (even away from the singularity ${y=-\infty}$). In particular, a peculiar behaviour is obtained for ${u=0}$, because ${\Delta_\e}$ depends on ${u}$ through the combination
\be
 z_\e=\frac{1}{\sqrt{2}}(\e^{-1}u+\e v) .
 \label{shortcut}
\ee
In order to have control over the exact distributional structure of the limit, it is convenient to isolate such dependence on $\e^{-1}u$ by performing first an expansion of ${\Delta_\e}$ with ${z_\e}$ unexpanded. This leads to an expression 
\be
 \Delta_\e=\frac{1}{\e}h(z_\e)\d u^2+\left[k_1(z_\e)\d x_2+k_2(z_\e)\d y_2+k_3(z_\e)\d z_2+k_4(z_\e)\d u\right]
        \d u+\dots ,
 \label{dominant}
\ee
where the dots denote terms proportional to higher powers of ${\e}$, which are negligible in the limit. We have emphasized here the dependence of the functions $h$ and ${k_i}$ ($i=1, \ldots, 4$) on $z_\e$
(and thus on $\e$), because this is essential in our limit, but they depend also on $x_2$, $y_2$ and $z_2$. The quantities ${k_i}$ are rather involved, but it suffices to observe here that $\lim_{\e\to 0}k_i(z_\e)=0$. We can thus also drop all the terms of order ${\e^0}$ in (\ref{dominant}).\footnote{\label{note-rescaling}A remark on the ``triviality'' of the ${\e^0}$ terms is in order, since they could be non-vanishing for certain more general scalings of the original metric parameters. While with higher order (in $\e$) corrections in Eq.~(\ref{ASlambda}) $\lim_{\e\to 0}k_i(z_\e)=0$ would still hold, we could introduce a non-vanishing contribution by allowing an $\e$-dependence in the ring ``radius'' via ${L_\e=L+c_1\e+c_2\e^2+\ldots}$. The convergence of the integral~(\ref{integ}) would then require $c_1=0$, but the quantity $c_2\e^2$ would affect the limit of~(\ref{dominant}) via ${\lim_{\e\to 0}k_4(z_\e)=c_2}$. The resulting term $c_2\d u^2$ is, however, obviously removable with a coordinate transformation.} 
For ${h}$, after all the steps described above, we obtain explicitly 
\beqn
 h(z_\e)= & & \P\frac{L^2}{\Sigma}+\Q\frac{L^2}{\Sigma^3}\left[(\xi^2-\eta^2-L^2)\frac{y_{1}}{\xi}\sin\alpha+2\xi x_{1}\cos\alpha\right]^2\nonumber \label{h_general} \\
  & & {}+\frac{1}{2}(2\Q-\P)\left(1-\frac{\xi^2+\eta^2-L^2}{\Sigma}\right)
\left(\frac{y_2^2}{\xi^2}\sin^2\alpha+\frac{x_2^2}{\eta^2}\cos\alpha\right) \nonumber \\
 & & {}+\sqrt{\P(\P-\Q)}\frac{Ly_2\sin\alpha}{\xi^2}\left(-1+\frac{\xi^2+\eta^2+L^2}{\Sigma}\right)
      +(\P-\Q)\frac{L^2y_2^2}{\xi^2\Sigma}\sin^2\alpha  \nonumber \\
 & & {}+\frac{1}{2}(\P-\Q)\left(1-\frac{\xi^2+\eta^2-L^2}{\Sigma}\right) .
\eeqn
Recall that the dependence of $h$ on $\e$ is contained in $x_1$ and $y_1$ via
Eqs.~(\ref{rotated})--(\ref{lorentzboost}), in $\eta$ and $\xi$ via
Eq.~(\ref{cartesian}) and in $\Sigma$ via Eq.~(\ref{Sigma}). In taking the
limit $\e\to 0$ of Eq.~(\ref{dominant}), we apply the distributional identity
\be
 \lim_{\e\to 0} \frac{1}{\e}f\left(z_\e \right)
  =\sqrt{2}\,\delta(u)\int_{-\infty}^{+\infty}f(z)\d z .
 \label{identity}
\ee
The final metric is thus [cf.~Eqs.~(\ref{back}) and (\ref{dominant})]
\be
 \d s^2=2\d u\d v+\d x_2^2+\d y_2^2+\d z_2^2+H(x_2,y_2,z_2)\delta(u)\d u^2 ,
 \label{ppgeneral}
\ee with a profile function given by \be
 H(x_2,y_2,z_2)=\sqrt{2}\int_{-\infty}^{+\infty}h(z)\d z .
 \label{integ}
\ee A black ring boosted to the speed of light in a general direction $z_1$ is
thus described by the metric~(\ref{ppgeneral}) with Eq.~(\ref{integ}). This is evidently a ${D=5}$ impulsive \pp wave with wave vector $\pa_v$. Such a spacetime is flat everywhere except on the null hyperplane $u=0$, which represents the impulsive wave front. Note that the equilibrium condition~(\ref{equilibrium2})
has not yet been enforced in the above expression for $h$ (in particular, in
the static limit $p_\nu=p_\lambda$ we recover the result of
\cite{OrtKrtPod05}). In order to write the solutions in a completely explicit
form, it only remains to perform the integration in Eq.~(\ref{integ}), with $h$
given by Eq.~(\ref{h_general}) with Eqs.~(\ref{Sigma})--(\ref{lorentzboost})
and~(\ref{shortcut}). For any $\alpha$, this integral is always convergent and
can in principle be expressed using elliptic integrals (because $\Sigma$ is a
square root of a fourth order polynomial in $z$, see \cite{OrtKrtPod05} for
related comments). Therefore, no singular coordinate transformation of the type
of \cite{AicSex71} has to be performed. In the following, we will explicitly calculate the integral, and study the
corresponding solution in the case of two different boosts of the black ring
along the privileged axes $x_1$ ($\alpha=0$) and $y_1$ ($\alpha=\pi/2$), which
are respectively ``orthogonal'' and ``parallel'' to the 2-plane $(y_1,y_2)$
[i.e., $(\xi,\psi)$] in which the ring rotates.

\section{Orthogonal boost: $\alpha=0$}

\label{sec-orthogonal}

For the orthogonal boost $\alpha=0$, from Eq.~(\ref{rotated}) one has $z_1=x_1$ and $z_2=y_1$, so that the general \pp wave~(\ref{ppgeneral}) reduces to
\be
 \d s^2=2\d u\d v+\d x_2^2+\d y_1^2+\d y_2^2+H_{_{\!\bot}}\!(x_2,y_1,y_2)\delta(u)\d u^2 .  
 \label{pporth}
\ee
Also, it is now convenient to rewrite $h$ in Eq.~(\ref{h_general}) as
\beqn
 h_{_{\!\bot}}\!(z_\e)= & & \left[3\P L^2-(\P-\Q)\xi^2-\Q(x_2^2+L^2)\right]\frac{1}{2\Sigma}+\Q\frac{4L^2\xi^2z^2_\e }{\Sigma^3} \nonumber \label{h_orth} \\
    & & {}+\frac{1}{2}(2\Q-\P)\left[\frac{x_2^2(L^2-\xi^2)}{(z^2_\e +x_2^2)\Sigma}+\frac{x_2^2}{z^2_\e   
          +x_2^2}\right]+\frac{1}{2}(\P-\Q)\left(1-\frac{z_\e^2}{\Sigma}\right) , 
\eeqn
and $\Sigma$ [from Eq.~(\ref{Sigma})] as 
\be
 \Sigma=\sqrt{\left[z_\e^2+x_2^2+(\xi+L)^2\right]\left[z_\e^2+x_2^2+(\xi-L)^2\right]} .
 \label{Sigma_orth}
\ee
Hereafter, it is understood that $\xi=\sqrt{y_1^2+y_2^2}$. In the orthogonal boost there is no contribution to $h_{_{\!\bot}}$ from the off-diagonal term $g_{t\psi}$ in the metric~(\ref{ring}). 
Performing the integration~(\ref{integ}) with $h$ given by Eqs.~(\ref{h_orth}) and (\ref{Sigma_orth}), we find
\beqn
 & & H_{_{\!\bot}}\!(x_2,y_1,y_2)=\sqrt{2}\,\frac{3\P L^2+(2\Q-\P)\xi^2}{\sqrt{(\xi+L)^2+x_2^2}}\,K(k)+\sqrt{2}(2\Q-\P) \nonumber   \label{Horth} \\ 
 & & \hspace{.15cm} {}\times \left[-\sqrt{(\xi+L)^2+x_2^2}\,E(k)+
      \frac{\xi-L}{\xi+L}\frac{x_2^2}{\sqrt{(\xi+L)^2+x_2^2}}\,\Pi(\rho,k)+\pi|x_2|\Theta(L-\xi)\right] , 
\eeqn
where
\be
 k=\sqrt{\frac{4\xi L}{(\xi+L)^2+x_2^2}} , \qquad \rho=\frac{4\xi L}{(\xi+L)^2} ,
 \label{krho_orth}
\ee
and $\Theta(L-\xi)$ denotes the step function. In the above calculation, we have used the standard elliptic integrals and their properties summarized in the Appendix of \cite{OrtKrtPod05}, and the additional integral [$\Sigma$ given by Eq.~(\ref{Sigma_orth}) with $z_\e$ replaced by $z$] 
\be
 \int_0^\infty\left(1-\frac{z^2}{\Sigma}\right)\d z=\sqrt{(\xi+L)^2+x_2^2}\,E(k) .
\ee

In order to gain physical insight, it is useful to visualize the behaviour of the gravitational field at a large spatial distance within the wave front $u=0$. Defining the coordinates~$(r,\theta)$
\be
 x_2=r\cos\theta , \qquad \xi=r\sin\theta ,
\ee
an expansion for small values of the dimensionless parameter $L/r$ (using the identities summarized in \cite{OrtKrtPod05}) leads to
\beqn
 H_{_{\!\bot}}= & & \frac{\pi}{\sqrt{2}}\P L\left[3\frac{L}{r}-\left(\frac{5}{8}+\frac{\Q}{4\P}\right)(3\cos^2\theta-1)\frac{L^3}{r^3}\right. \nonumber  \label{multipole_orht} \\
    & & {}+\left.\left(\frac{7}{64}+\frac{\Q}{16\P}\right)(35\cos^4\theta-30\cos^2\theta+3)\frac{L^5}{r^5}
      +O\left(\frac{L^7}{r^7}\right)\right] .
\eeqn
We recognize the standard form of multipole terms. The monopole is essentially an \AS term. The dipole and the octupole are missing, due to the geometry of the source. The quadrupole and 16-pole reflect the shape of the singularity and depend on the spin of the original black ring, but they persist even in the static limit $\Q=\P$ \cite{OrtKrtPod05} (when, in fact, they reach their maximal strength).

It is remarkable that for the physically more interesting case of {\em black rings in equilibrium}, i.e. those satisfying $\P=2\Q$ [see Eq.~(\ref{equilibrium2})], the profile function simplifies significantly to
\be
 H^e_{_{\!\bot}}(x_2,y_1,y_2)=\frac{3\sqrt{2}\,\P L^2}{\sqrt{(\xi+L)^2+x_2^2}}\,K(k)
 \label{Horth_eq} .
\ee
Interestingly, this is just the Newtonian potential generated by a {\em uniform ring of radius $L$} and linear density $\mu=3\sqrt{2}\P L/4$ located at $x_2=0$ in the flat three-dimensional space $(x_2,y_1,y_2)$. Since for a general \pp wave~(\ref{pporth}) the only component of the Ricci tensor is ${R_{uu}=-\frac{1}{2}\delta(u){\bf\Delta}H_{_{\!\bot}}}$, ${\bf\Delta}$ denoting the Laplace operator over the transverse space $(x_2,y_1,y_2)$, it follows that the profile function~(\ref{Horth_eq}) represents a spacetime which is vacuum everywhere except on the circle $u=0=x_2$, $\xi=L$ [so that $k=1$ in Eq.~(\ref{krho_orth})]. This lies on the wave front and corresponds to a singular ring-shaped source moving with the speed of light. It is obviously a remnant of the curvature singularity ($y=-\infty$) of the original static black ring (\ref{ring}). For the non-equilibrium solution~(\ref{Horth}), the discontinuous term proportional to $\Theta(L-\xi)$ is responsible for a disk memebrane supporting the ring \cite{OrtKrtPod05}. We have plotted typical profile functions $H_{_{\!\bot}}$ and $H^e_{_{\!\bot}}$ in Fig.~\ref{fig_orth}. 
\FIGURE[ht]{
 \epsfig{file=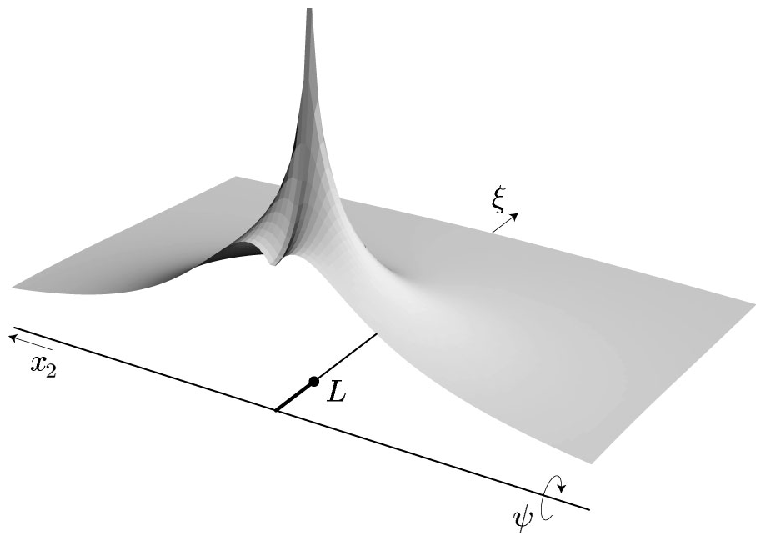}\hspace*{10pt}
 \epsfig{file=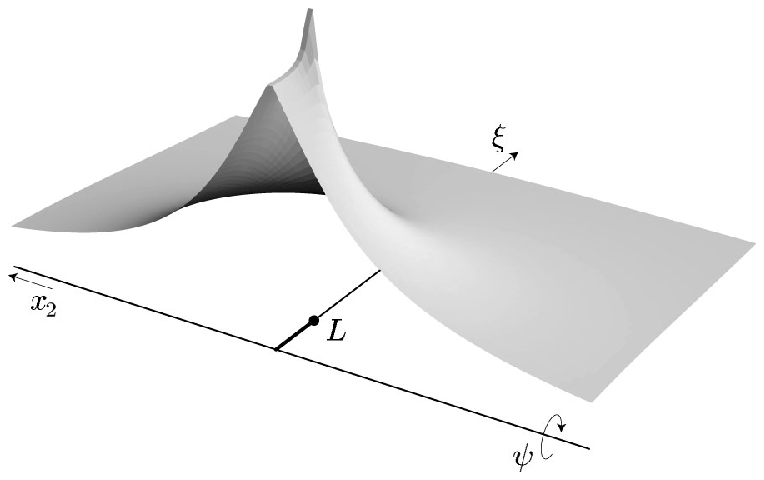}\hspace*{10pt} \\
 \epsfig{file=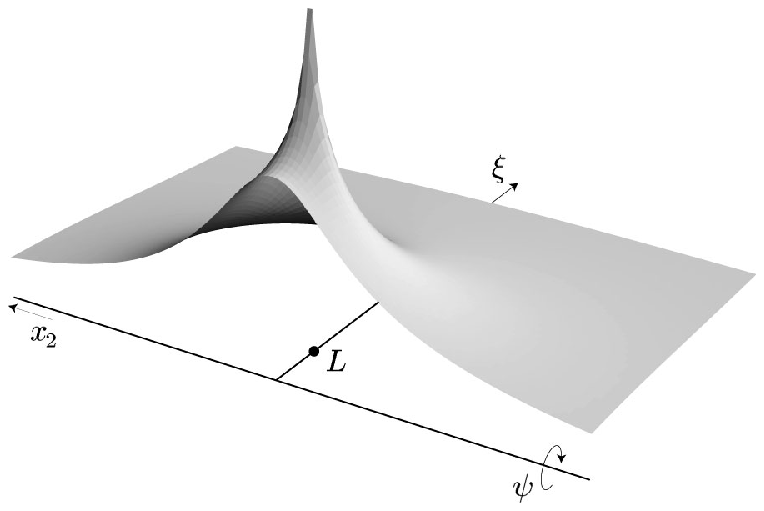}
\caption{The profile function $H_{_{\!\bot}}$, given by Eq.~(\ref{Horth}), in the case of underspinning ($\P<2\Q$, left), overspinning ($\P>2\Q$, right) and balanced ($\P=2\Q$, bottom) black rings [cf. Eq.~(\ref{equilibrium2})] boosted along an orthogonal direction $x_1$. It is represented over the plane $(x_2,\xi)$ [cf.~Eq.~(\ref{cartesian})], and the Killing coordinate $\psi$ is suppressed. In the equilibrium case, $H_{_{\!\bot}}$ reduces to $H^e_{_{\!\bot}}$ of Eq.~(\ref{Horth_eq}) and the disk membrane at $x_2=0$, $\xi<L$ disappears (no jump of $\pa H^e_{_{\!\bot}}/\pa x_2$ occurs at $x_2=0$). In all cases, there is a ring singularity at ${x_2=0}$, ${\xi=L}$, as indicated by the thick points in the pictures.}
\label{fig_orth}}

\section{Parallel boost: $\alpha=\pi/2$}

\label{sec-parallel}

For the parallel boost $\alpha=\pi/2$, from Eq.~(\ref{rotated}) one has $z_1=y_1$ and $z_2=-x_1$, and the general \pp wave~(\ref{ppgeneral}) reduces to
\be
 \d s^2=2\d u\d v+\d x_1^2+\d x_2^2+\d y_2^2+H_{_{\!||}}\!(x_1,x_2,y_2)\delta(u)\d u^2 .  
 \label{ppparall}
\ee
The function $h$ can be reexpressed as
\beqn
 h_{_{||}}\!(z_\e)= & & \left[(3\P+\Q)L^2-\Q y_2^2+2\sqrt{\P(\P-\Q)}Ly_2-(\P-\Q)\eta^2\right]\frac{1}{2\Sigma }-\Q\frac{4L^2\eta^2z_\e^2}{\Sigma^3} \nonumber \label{h_parall} \\
 & & {}+\frac{1}{2}\left[(2\Q-\P)y_2^2-2\sqrt{\P(\P-\Q)}Ly_2\right]\left[-\frac{L^2+\eta^2}{(z_\e^2 +y_2^2)\Sigma}+\frac{1}{z^2_\e+y_2^2}\right] \nonumber \\ 
 & & {}+\frac{1}{2}(\P-\Q)\left(1-\frac{z_\e^2}{\Sigma}\right) , 
\eeqn
and
\be
 \Sigma=\sqrt{z_\e^4+2(y_2^2+\eta^2-L^2)z_\e^2+a^4} , \label{Sigma_parall}
\ee
with
\be
 a=\left[(\eta^2+y_2^2-L^2)^2+4\eta^2L^2\right]^{1/4} . 
 \label{def_a}
\ee
It is understood that ${\eta=\sqrt{x_1^2+x_2^2}}$.
\FIGURE[ht]{
 \epsfig{file=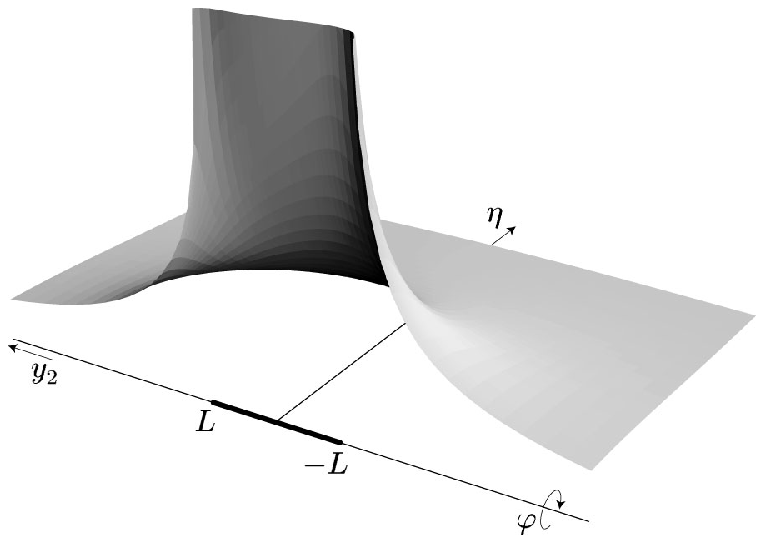}\hspace*{10pt}
 \epsfig{file=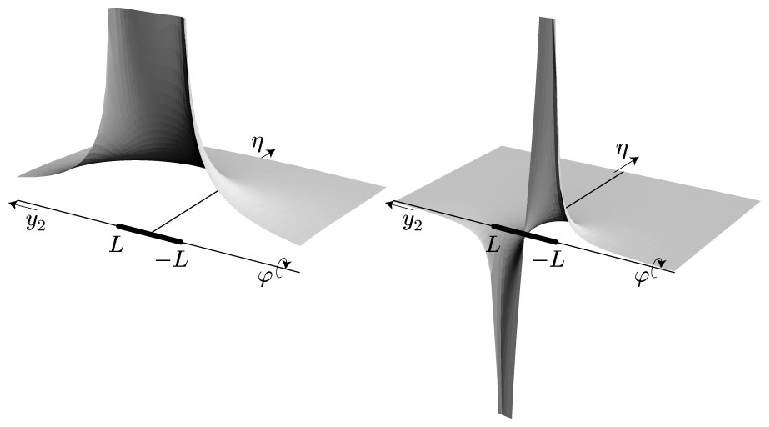}
\caption{Plot of the profile function $H^e_{_{\!||}}(y_2,\eta)$, given by Eq.~(\ref{Hparall_eq}), for balanced black rings ($\P=2\Q$) boosted along the direction $y_1$ in the plane of rotation. It is depicted over the plane $(y_2,\xi)$ [cf.~Eq.~(\ref{cartesian})], and the Killing coordinate $\phi$ is suppressed. The profile function $H^e_{_{\!||}}$ diverges at the rod singularity $\eta=0$, $|y_2|\le L$, as indicated by the thick line. The two smaller pictures represent the symmetric and antisymmetric part (with respect to the origin of the $y_2$-axis) of $H^e_{_{\!||}}$, respectively. The case of unbalanced black rings, Eq.~(\ref{Hparall}), does not produce qualitative changes, since the disk membrane Lorentz-contracts to the singular rod region.} 
 \label{fig_parall}}
Performing the integration~(\ref{integ}) with $h$ given by Eqs.~(\ref{h_parall})--(\ref{def_a}), one obtains
\beqn
 & & H_{_{\!||}}\!(x_1,x_2,y_2)= \left[2(2\P-\Q)L^2+(2\Q-\P)a^2\left(1+\frac{L^2+\eta^2}{a^2-y_2^2}\right)\right. \nonumber \\ & & \hspace{2.5cm} \left.+2\sqrt{\P(\P-\Q)}Ly_2\left(1-\frac{L^2+\eta^2}{a^2-y_2^2}\right)\right]
\frac{\sqrt{2}}{a}K(k)-2\sqrt{2}(2\Q-\P)aE(k) \nonumber \label{Hparall} \\ 
 & & \quad                
 \hspace{-.3cm}{}+\frac{\sqrt{2}}{2}\left[(2\Q-\P)y_2-2\sqrt{\P(\P-\Q)}L\right]\left[-\frac{\eta^2+L^2}{ay_2}\frac{a^2+y_2^2}{a^2-y_2^2}\,
    \Pi(\rho,k)+\pi\,\mbox{sgn}(y_2)\right] , 
\eeqn
where
\be
 k=\frac{\left(a^2-\eta^2-y_2^2+L^2\right)^{1/2}}{\sqrt{2}a} , \qquad     
      \rho=-\frac{(a^2-y_2^2)^2}{4a^2y_2^2} .
 \label{krho_parall}     
\ee
Again, we refer to the Appendix of \cite{OrtKrtPod05}, the only additional integral used here being [$\Sigma$ given by Eq.~(\ref{Sigma_parall}) with $z_\e$ replaced by $z$]
\be
 \int_0^\infty\left(1-\frac{z^2}{\Sigma}\right)\d z=2aE(k)-aK(k) .
\ee

With the coordinates
\be
 y_2=r\cos\theta , \qquad \eta=r\sin\theta ,
\ee
the behaviour at large spatial distances is given by
\beqn
 H_{_{||}}= & & \frac{\pi}{\sqrt{2}}\P L\left[3\frac{L}{r}+2\sqrt{\frac{\P-\Q}{\P}}\cos\theta\,\frac{L^2}{r^2}+\left(\frac{7}{8}-\frac{\Q}{4\P}\right)(3\cos^2\theta-1)\frac{L^3}{r^3}
      \right.\nonumber  \label{multipole_parall} \\ 
      & & {}+\left.\frac{3}{4}\sqrt{\frac{\P-\Q}{\P}}(5\cos^3\theta-3\cos\theta)\frac{L^4}{r^4}\right. \nonumber \\ 
      & & {}+\left.\left(\frac{11}{64}-
        \frac{\Q}{16\P}\right)(35\cos^4\theta-30\cos^2\theta+3)\frac{L^5}{r^5}+O\left(\frac{L^6}{r^6}\right)\right] .
\eeqn
Notice that now there appear also a dipole and an octupole term, as a remnant of the angular momentum of the black ring.

We are especially interested in {\em black rings in equilibrium}~(\ref{equilibrium2}), for which one is left with
\beqn
 H^e_{_{\!||}}(x_1,x_2,y_2)= & & \P L \left[\frac{3\sqrt{2}L}{a}+\frac{2y_2}{a}\left(1-\frac{L^2+\eta^2}{a^2-y_2^2}\right)\right]K(k) \nonumber \label{Hparall_eq} \\  & & {}+\P L\left[\frac{\eta^2+L^2}{ay_2}\frac{a^2+y_2^2}{a^2-y_2^2}\,\Pi(\rho,k)
    -\pi\,\mbox{sgn}(y_2)\right] . 
\eeqn
This function is singular at the points satisfying $u=0=\eta$ and $|y_2|\le L$ [$k=1$ in Eq.~(\ref{krho_parall})], i.e. on a {\em rod of length $2L$} contained within the wave front. This is a remnant of the curvature singularity of the original static black ring (\ref{ring}), which has (infinitely) Lorentz-contracted because of the ultrarelativistic boost in the plane of the ring. For the same reason, and because the original ring was rotating, the rod-source corresponding to Eq.~(\ref{Hparall_eq}) is not uniform. The profile~(\ref{Hparall_eq}) corresponds to a vacuum spacetime everywhere except on the rod. Notice also that the apparent divergences of $H^e_{_{\!||}}$ at $y_2^2=a^2$ and $y_2=0$ is only a fictitious effect: the singular behaviour of the coefficient of $\Pi$ in Eq.~(\ref{Hparall_eq}) is exactly compensated from that of $K$ in the first case and from the $\mbox{sgn}(y_2)$ function in the second case [recall also the form of $\rho$ in Eq.~(\ref{krho_parall})]. Finally, it is interesting to observe that the antisymmetric part (in the coordinate $y_2$) of $H_{_{\!||}}$ and $H^e_{_{\!||}}$ comes entirely from the off-diagonal term $g_{t\psi}$ in the metric~(\ref{ring}), which was responsible for rotation before the boost [and produces the terms even in $\cos\theta$ in the expansion~(\ref{multipole_parall})]. The profile function $H^e_{_{\!||}}$ is plotted in Fig.~\ref{fig_parall}. 

\section{Boost of the supersymmetric black ring}

To conclude, we demonstrate that the above method can also be employed to calculate the gravitational field generated by other black rings in the ultrarelativistic limit. The first supersymmetric black ring (solution of $D=5$ minimal supergravity) was presented in \cite{Elvangetal04} (and subsequently generalized in \cite{Elvangetal05,BenWar04,GauGut05}). The line element reads
\be
 \d s^2=-f^2(\d t+\omega_\psi\d\psi+\omega_\phi\d\phi)^2+f^{-1}(\d s_0^2+\d t^2) , 
 \label{supersym}
\ee
with $\d s_0^2$ as in Eq.~(\ref{background}) and 
\beqn
 f^{-1} & = & 1+\frac{Q-q^2}{2L^2}(x-y)-\frac{q^2}{4L^2}(x^2-y^2) , \\
 \omega_\psi & = & \frac{3}{2}q(1+y)+\frac{q}{8L^2}(1-y^2)\left[3Q-q^2(3+x+y) \right] , \\
 \omega_\phi & = & -\frac{q}{8L^2}(1-x^2)\left[3Q-q^2(3+x+y)\right] .
\eeqn
The $S^1\times S^2$ horizon is localized at $y\to-\infty$, and asymptotic infinity at $x,y\to -1$. The Maxwell field $F=\d A$ is determined by
\be
 A=\frac{\sqrt{3}}{2}f(\d t+\omega_\psi\d\psi+\omega_\phi\d\phi)-\frac{\sqrt{3}}{4}q[(1+x)\d\phi+(1+y)\d\psi] .
\ee
The net electric charge and the local dipole magnetic charge are proportional to the positive parameters $Q$ and $q$, respectively, which (for a physical interpretation) are assumed to satisfy $Q\ge q^2$ and $L<(Q-q^2)/(2q)$ \cite{Elvangetal04}. The mass and angular momenta of the ring are
\be
 M=\frac{3\pi}{4}Q , \qquad J_\psi=\frac{\pi}{8}q(6L^2+3Q-q^2) , \qquad J_\phi=\frac{\pi}{8}q(3Q-q^2) .
 \label{charge_super}
\ee
In the limit $q=0$ the black ring becomes a static charged naked singularity, solution of the pure Einstein-Maxwell theory. In order to boost the line element~(\ref{supersym}), we can follow a procedure almost identical to the one used for the vacuum ring. The standard mass rescaling of \cite{AicSex71} together with the inequality $L<(Q-q^2)/(2q)$ suggests that during the boost we rescale the charges as
\be
 Q=\e p_Q , \qquad q=\epsilon p_q  \qquad (p_Q>2Lp_q).
 \label{rescaling_super}
\ee
Omitting straightforward intermediate steps, in the case of a boost orthogonal to the plane $(\xi,\psi)$ we obtain a shock \pp wave~(\ref{pporth}) with 
\be
 H^s_{_{\!\bot}}(x_2,y_1,y_2)=\frac{3\sqrt{2}\,p_Q}{\sqrt{(\xi+L)^2+x_2^2}}\,K(k)
 \label{Horth_super} ,
\ee
and $k$ given by Eq.~(\ref{krho_orth}). For a parallel boost, we obtain the metric~(\ref{ppparall}) with 
\beqn
 H^s_{_{\!||}}(x_1,x_2,y_2)= & & 3\sqrt{2} \left[p_Q\frac{1}{a}+p_q\frac{y_2}{a}\left(1-\frac{L^2+\eta^2}{a^2-y_2^2}\right)\right]K(k) \nonumber \label{Hparall_super} \\  & & {}+\frac{3\sqrt{2}p_q}{2}\left[\frac{\eta^2+L^2}{ay_2}\frac{a^2+y_2^2}{a^2-y_2^2}\,\Pi(\rho,k)
    -\pi\,\mbox{sgn}(y_2)\right] ,
\eeqn
where $k$ and $\rho$ as in Eq.~(\ref{krho_parall}). To obtain the field of a boosted naked singularity ($q=0$) just set $p_q=0$ in Eq.~(\ref{Hparall_super}). Notice that the dipole charge $q$ has an effect only in the case of a parallel boost, since $p_q$ does not appear in $H^s_{_{\!\bot}}$ [which is in fact equivalent to the expression~(\ref{Horth_eq}) for balanced vacuum rings]. This is related to the ``asymmetry'' between the angular momenta $J_\psi$ and $J_\phi$ in Eq.~(\ref{charge_super}). 
In both boosts, one also finds that ${F=\d A}$ tends to zero together with its associated energy-momentum tensor (so that the ``peculiar configuration'' of \cite{LouSan90} does not arise here). In fact, both $H^s_{_{\!\bot}}$ and $H^s_{_{\!||}}$ correspond to vacuum \pp waves. In principle, rescalings different from Eq.~(\ref{rescaling_super}) can be considered if one drops the requirement $L<(Q-q^2)/(2q)$. The detailed investigation of this and other possibilities is left for possible future work.

\acknowledgments

M.O. is supported by a post-doctoral fellowship from Istituto Nazionale di Fisica Nucleare (bando n.10068/03).

\appendix

\section{Results for the boosted $D=5$ Myers-Perry black hole}

\label{app_yoshino}

Ref.~\cite{Yoshino05} analyzed the ultrarelativistic boost of $D$-dimensional Myers-Perry black holes \cite{MyePer86} with a single non-vanishing angular momentum. As in the present work, the calculation was performed in the case of two particular boosts orthogonal and parallel to the plane of rotation, and for $D=5$ it resulted in impulsive \pp waves of the type~(\ref{pporth}) and (\ref{ppparall}), respectively. It is thus interesting to compare the results of~\cite{Yoshino05} to ours. First of all, the angular momentum of black holes in $D=5$ must obey a Kerr-like bound $a^2<\mu$ \cite{MyePer86}. Since, in the \AS limit, Ref.~\cite{Yoshino05} sent the mass parameter $\mu$ to zero while keeping the spin parameter $a$ fixed, for $D=5$ the final metrics refer to boosted naked singularities rather than black holes \cite{Yoshino05}. On the other hand, there is no upper limit on the spin of black rings \cite{EmpRea02prl}, so that in our limit the rings do remain ``black'' until the final \pp wave is obtained (the same applies to the solutions of \cite{Yoshino05} in $D\ge 6$, when also black holes can be ultra-spinning). In the rest of this appendix we shall present the profile functions of \cite{Yoshino05} (for the case $D=5$) using an explicit form adapted to our notation\footnote{In particular, the quantity $L$ will replace the original spin parameter $a$.}, and we shall compare them with our functions (\ref{Horth}) and (\ref{Hparall}). 

\subsection{Orthogonal boost}

For an orthogonal boost, the result of \cite{Yoshino05} can be rearranged as
\beqn
 \tilde H_{_{\!\bot}}\!(x_2,y_1,y_2)= & & \frac{8\sqrt{2}\,p_M}{3\pi}\left[\frac{2\sqrt{2}}{(\xi^2+x_2^2+L^2+b^2)^{1/2}}\,K(k_1)+\frac{\sqrt{2}}{L^2}
          (\xi^2+x_2^2+L^2+b^2)^{1/2} \,E(k_1)\right.\nonumber   \label{HorthBH} \\ 
 & & {}-\left.\frac{2\sqrt{2}}{L^2}\frac{b^2}{(\xi^2+x_2^2+L^2+b^2)^{1/2}}\,\Pi(\rho_1,k_1)\right] ,
\eeqn
where
\beqn
 k_1 & = & \left(\frac{\xi^2+x_2^2+L^2-b^2}{\xi^2+x_2^2+L^2+b^2} \right)^{1/2} , \qquad \rho_1=-\frac{(\xi^2+x_2^2-L^2-b^2)^2}{4L^2x_2^2} , \nonumber  \label{krho_orthBH} \\ [5pt]
 b & = & [(\xi^2+x_2^2-L^2)^2+4x_2^2L^2]^{1/4} .
\eeqn
The above elliptic functions are singular for $k_1=1$, that is on a circle of radius $L$ given by $x_2=0$, $\xi=L$. This was already remarked in \cite{Yoshino05} and it resembles our results of Sec.~\ref{sec-orthogonal}. Other physical properties are more ``hidden'' in the expression~(\ref{HorthBH}). First of all, for $x_2\to 0$ one has $\rho_1\to 0$ if $\xi>L$, whereas $\rho_1$ diverges if $\xi<L$. This implies [with identity~(A5) of \cite{OrtKrtPod05}] that, when $\xi<L$ and $x_2$ is small, $H_{_{\!\bot}}$ contains a non-smooth term proportional to $|x_2|$, namely there is an additional membrane at $x_2=0$ and $\xi<L$ (i.e. within the ring singularity discussed above). The presence of such a disk-shaped source is related to the structure of the singularities of the Myers-Perry solutions \cite{MyePer86}, and it should be contrasted with the simpler profile function~(\ref{Horth_eq}) for balanced black rings, which has only a ``uniform'' circle as a source.  
From a complementary viewpoint, we can compare an expansion of the profile function~(\ref{HorthBH}) at large spatial distances with the analogous result~(\ref{multipole_orht}) for the black ring. From Eq.~(\ref{HorthBH}) we obtain
\be
 \tilde H_{_{\!\bot}}=\frac{1}{\sqrt{2}}\frac{8\,p_M}{3L}\left[3\frac{L}{r}-\frac{5}{8}(3\cos^2\theta-1)\frac{L^3}{r^3}+
      \frac{7}{64}(35\cos^4\theta-30\cos^2\theta+3)\frac{L^5}{r^5}+O\left(\frac{L^7}{r^7}\right)\right] .
 \label{multipole_orhtBH}
\ee
The monopole term coincides with the one in the corresponding expression~(\ref{multipole_orht}) for the black ring, which we should expect since we are boosting objects with the same mass (which scales as $M=\gamma^{-1}p_M$). However, Eqs.~(\ref{multipole_orhtBH}) and (\ref{multipole_orht}) in general differ already in the quadrupole term, in particular for the physically most interesting case of balanced rings $\P=2\Q$. They coincide only in the limiting case $\Q=0$, corresponding to $\nu=0$, when the black ring in fact reduces to a naked singularity isometric to that of Myers and Perry (see the discussion above about the Kerr bound). In addition, in the limit of vanishing rotation $L=0$ of Eq.~(\ref{multipole_orhtBH}) only the \AS monopole survives, which corresponds to the ultrarelativistic boost of the $D=5$ Schwarzschild-Tangherlini black hole.\footnote{Recall that, instead, balanced black rings can not be static, while unbalanced static rings correspond to setting $\Q=\P$ in Eq.~(\ref{multipole_orht}) \cite{OrtKrtPod05} (and {\em not} $L=0$).}

\subsection{Parallel boost}

For a parallel boost, the profile function of \cite{Yoshino05} is
\beqn
 \tilde H_{_{\!||}}\!(x_1,x_2,y_2)= & & \frac{8\sqrt{2}\,p_M}{3\pi}\left[\frac{4}{a}\left(1+\frac{\eta^2+y_2^2+L^2+a^2}{2Ly_2}\right)K(k)
+\frac{2a}{L^2}\,E(k)\right.\nonumber   \label{HparallBH} \\ 
 & & {}-\left.\frac{2L+y_2}{a}\frac{\eta^2+y_2^2+L^2+a^2}{L^2y_2}\,\Pi(\rho_1,k)\right] ,
\eeqn
with $a$ as in Eq.~(\ref{def_a}), $k$ as in Eq.~(\ref{krho_parall}) and
\be
 \rho_1=-\frac{\eta^2+y_2^2+L^2-a^2}{2a^2} .
 \label{rho_parallBH}     
\ee
Similarly as in Sec.~\ref{sec-parallel}, the elliptic integrals are singular at $k=1$, i.e. on a rod of length $2L$ located at $\eta=0$, $|y_2|\le L$ \cite{Yoshino05}. At large spatial distances, the expression~(\ref{HparallBH}) behaves as
\beqn
 \tilde H_{_{||}}= & & \frac{1}{\sqrt{2}}\frac{8\,p_M}{3L}\left[3\frac{L}{r}+2\cos\theta\,\frac{L^2}{r^2}+\frac{7}{8}(3\cos^2\theta-1)\frac{L^3}{r^3}
        +\frac{3}{4}(5\cos^3\theta-3\cos\theta)\frac{L^4}{r^4}\right.\nonumber  \label{multipole_parallBH} \\ 
      & & {}+\left.\frac{11}{64}(35\cos^4\theta-30\cos^2\theta+3)\frac{L^5}{r^5}+O\left(\frac{L^6}{r^6}\right)\right] .
\eeqn
The discussion is similar as the one above for $\tilde H_{_{\!\bot}}$. Again, the monopole term coincides with the one in the corresponding expression~(\ref{multipole_parall}) for the black ring. Higher multipoles in general differ, in particular for balanced rings. Boosted black holes reduce to the \AS monopole in the static limit $L=0$.



\providecommand{\href}[2]{#2}\begingroup\raggedright\endgroup

\end{document}